\documentclass[prl,twocolumn,showpacs, amsmath,amssymb,aps]{revtex4-1}
\usepackage{graphicx}
\usepackage{bm}

\begin{document}
\newcommand{\balpha}{\mbox{\boldmath$\alpha$}}

\title{Saddle point singularity and optical phase transition in bilayer hyperbolic metamaterials}

\author{Vitaliy N. Pustovit, David E. Zelmon, Kurt G. Eyink and Augustine M. Urbas}

\affiliation{Materials and Manufacturing Directorate, Air Force Research Laboratory, Wright Patterson Air Force Base, Ohio 45433, USA}


\begin{abstract}
 We study theoretically and numerically high density of states for hyperbolic bilayered metamaterials (HMM). It reveals that
 density response of HMM is reminiscent of Fermi electronic band structure of metal or semiconductors.
 By the method of Green function a van Hove type singularity is found in photonic density spectra of HMM with saddle point localization on photonic
 Fermi surface (FS) of metamaterial. Similar to the electronic systems, the photonic FS experiences instabilities induced by
 the changes in volume fractions of its constituents that leads to the Lifshitz type zero-temperature phase transition between FS of types I and II
 hyperbolic states at the topology protected critical point.
 \end{abstract}


\maketitle
 Properties of condensed matter materials are highly determined by their surfaces. One of the key parameters that determines a
 material properties is the electromagnetic (photonic) density of states (DOS).
 Over the last few decades there have been many attempts to exploit a variety of systems with high density of states \cite{stefano,cortes,jacob}.
 Microcavities and photonic
 crystals as well as plasmonic systems based on metal nanoparticles can be considered as the most promising candidates for producing high DOS.
 However, these approaches imply a set of restrictions on spectral width and resonance position which creates difficulties in their
 practical use. As promising alternative is an artificial metamaterials with a hyperbolic dispersion. This system
 could simultaneous excite a large number of electromagnetic states producing a broadband DOS enhancement.
 Recently it has been attempts to bring a concept of symmetry in topological order to the simple HMM systems \cite{narim1,narim-ishii} though a rigorous evidences of the non-trivial topological behavior were not established. Later, an appropriate justification of topological behavior of HMM was provided but for the more complex system with magnetic chirality \cite{zhang}. In this manuscript we discuss some symmetry aspects that nevertheless arise in the DOS behavior of the topologically trivial HMM systems.

  The change in Fermi surface in metals and semiconductors from an open topology to a closed one is directly related to singularities in
  the density of states of HMM (Van Hove singularities \cite{hove}). By the method of Green function we analyze the positions of these singularities and compare them to positions of phase transitions and photonic Fermi band structure. We found these transitions to those for the band structure of the metals or semiconductors.

\begin{figure}
 \centering
 \includegraphics[width=\columnwidth]{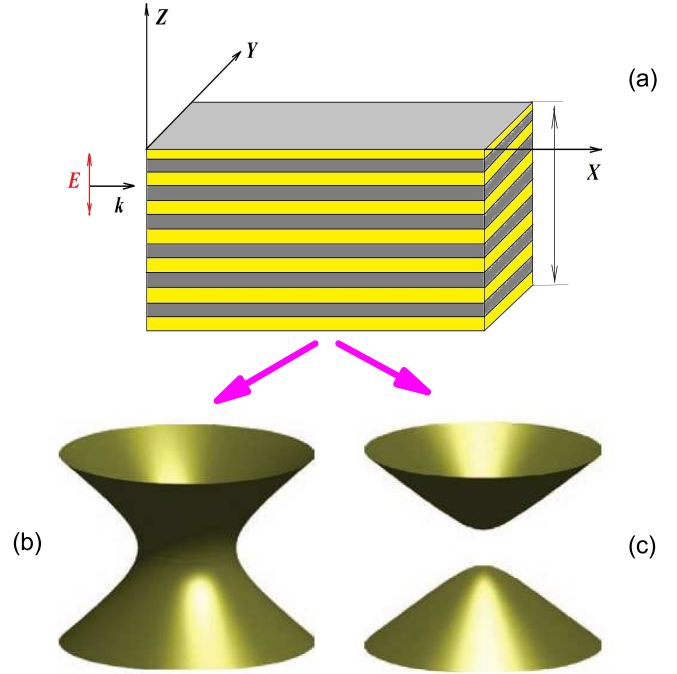}
 \caption{\label{fig:hmm} (color online) Multilayer HMM structure (a) with corresponding iso-frequency surfaces of "neck touched"  type II (b) when
 $\epsilon_\| >0$ and $\epsilon_\bot <0$ and "neck disrupted" of type I (c) when $\epsilon_\| <0$ and $\epsilon_\bot >0$.}
 \end{figure}

\begin{figure}
 \centering
 \includegraphics[width=\columnwidth]{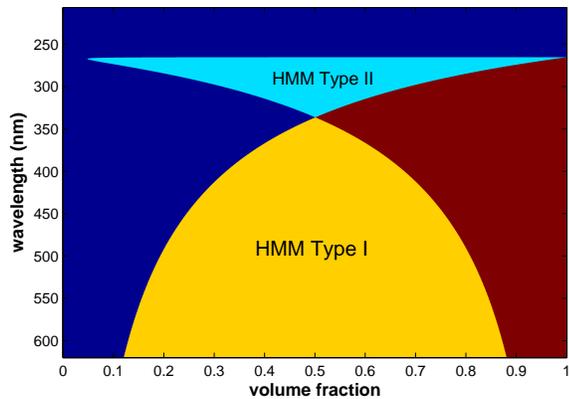}
 \caption{\label{fig:epsilon} (color online) Topologically protected state at $f=0.5$ for multilayer HMM with changing
 fraction of silver inside of stack.}
 \end{figure}

 We present a complete description of the spectral response of the composite HMM system using the example of
 a multilayer stack of silver and silica (see Fig.\ref{fig:hmm}).
The system is illuminated by an incident TM polarized wave, assuming propagation of plasmonic modes inside HMM along the x-axis.
Due to the nontrivial symmetry, the photon isofrequency surface inside of HMMs defined by the properties of an extraordinary (transverse magnetic polarized) wave, which is given by \cite{shalaev2}
\begin{eqnarray}
\label{dispersion}
\frac{k_x^2+k_y^2}{\epsilon_\parallel}+\frac{k_z^2}{\epsilon_\perp} = \frac{\omega^2}{c^2},
\end{eqnarray}
with having either $\epsilon_\| <0$ and $\epsilon_\bot >0$ for type I or  $\epsilon_\| >0$ and $\epsilon_\bot <0$
for type II as shown in Fig.\ref{fig:epsilon}.
Looking at the band structure of the FS (see Fig.\ref{fig:hmm} (b-c) and Fig.\ref{fig:epsilon}) one may notice similarities with the classical Lifshitz \cite{lifshitz} phase transitions that were initially found in conditions of high pressure and low temperature in the momentum space of FS of solids with a change in the chemical potential which is reminiscent of the change of volume fraction of metal. Indeed, the phase transitions between the different types of HMMs shown in Fig.\ref{fig:epsilon} can thought of as a Landau-like transitions. By definition these phase transitions occur between phases with different symmetry groups and have a trivial topology with zero Berry curvature and topological charge \cite{volovik}.
Though these phase transitions may possess topological character with the presence of non-zero topological charges, it generally occurs when time-reversal or inversion symmetry becomes broken which typically requires the presence of a magneto-optical material within the stack \cite{zhang}. However, within the model considered in this manuscript, we restrict ourselves only to the trivial Landau-type phase transitions in a HMM. It is well known that as local perturbations of the bulk cannot affect properties of the edge states, these edge states are said to be topologically protected \cite{wein}.  Our direct calculations demonstrate that transition point between hyperbolics of types I and II occurs at $f=0.5$ (see Fig.\ref{fig:epsilon}), as it follows directly from the solution of transition equation $\epsilon'_\parallel \cdot\epsilon'_\bot=0$, is robust against changes in the dielectric constants of the metallic or dielectric layers as long as hyperbolic state is preserved. That brings us to a comparison in the topological protection of the given state with 3D gapless electronic systems as topological insulator \cite{hughe}, where Dirac points are topologically protected by property of the momentum conservation. Here, for the photonic systems, the same mechanism of the momentum conservation leads to formation of the saddle point and the van Hove type singularities in the position of the Type I and Type II phase transition.

Our approach is based on the Green's function method to characterize the surrounding media, which will allow obtaining the enhanced density
of states of the HMM. Here, we are concerned mostly with near field interaction of an incident light with the surface of HMM given in form
of Sommerfeld integral \cite{sommer,maradudin,markel}
\begin{eqnarray}
\label{sommer}
{\bf \bar{G}}({\bf r},{\bf r}')=\int \frac{d^2 q}{\chi(q)} {\bf \bar{D}}(q) \exp[iq(\sigma-\sigma')-\chi(q)(z+z')]
\end{eqnarray}
where ${\bf r}=(\sigma,z)$ and ${\bf r'}=(\sigma',z')$ are corresponding coordinates of the source and observation points,
 $\sigma=\sqrt{x^2 + y^2}$, and $\chi(q)=\sqrt{q^2-k^2}$. The operator ${\bf \bar{D}}(q)$ is a function of the Jacobi rotation
 matrix and reflection coefficient \cite{markel}. Exact analytical evaluation of the Green function tensor can be very cumbersome.
  However, for certain geometries, the expression for the Green tensor can be significantly simplified. In particular, a suitable
  analytical expression can be obtained with incident TM field when emitter's dipole is oriented perpendicular to the surface of the hyperbolic
  metamaterial (see Appendix \cite{pust1}). It further can be simplified by performing volume averaging
  to obtain a suitable expression for the Green function inside of the metamaterial. It can be shown that Green function can be presented with
  short-distance expansion in powers of $(kL)$, where $L$ is a characteristic length, defined as the distance from the point of observation
  to the reflected position of the source with respect to the plain interface.
\begin{eqnarray}
\label{expansion}
G= \frac{1}{4\pi k^2 L^3} \sum_{l=0}^{\infty} (k_1 L)^l K^{(l)},
\end{eqnarray}
where zero order of expansion
\begin{eqnarray}
 K^{(0)}= R_{12} \left[3\frac{Z^2}{L^2} -1\right],
 \end{eqnarray}
would correspond to the non-radiative (non-emissive) part of the Green's function
\begin{eqnarray}
G_{zz}^{nr} = \frac{R_{12}}{4 \pi k^2 L^3} \left[3\frac{Z^2}{L^2} -1\right],
 \end{eqnarray}
and where the characteristic length $L$ reduces to $Z$ coordinate. Parameter $R_{12}$,
which can be interpreted as a reflection coefficient of p-polarized light in the limit of high k-waves \cite{narim1, sipe, shalaev},
was derived in \cite{pust1}
\begin{eqnarray}
\label{polariz}
R_{12}= \frac{\epsilon_x-\epsilon_0 \sqrt{\frac{\epsilon_x}{\epsilon_z}}}{\epsilon_x+\epsilon_0 \sqrt{\frac{\epsilon_x}{\epsilon_z}}},
\end{eqnarray}
 where $\epsilon_0=2.2$ is a dielectric constant of medium above HMM. The two independent components
 $\epsilon_\bot=\epsilon_x=\epsilon_y$ and $\epsilon_\| =\epsilon_z$ of the HMM uniaxial tensor are determined by Maxwell-Garnett formulas for effective medium
 approximation. While this approach generally overestimates value of the density of states \cite{sipe,sipe2,maria}, it serves as a first
 approximation for the purpose of this paper.
 For a multilayer HMM, components of dielectric uniaxial tensor can be found \cite{drachev} as
\begin{eqnarray}
\label{garnett}
\epsilon_x=\epsilon_y=f \epsilon_m + (1-f)\epsilon_d,
\nonumber\\
\epsilon_z=\left[f\epsilon^{-1}_m + (1-f) \epsilon^{-1}_d\right]^{-1},
\end{eqnarray}
where $\epsilon_m$ and $\epsilon_d$ are dielectric constants of the aluminium and silica, respectively \cite{christy} and $f$ is
the metal filling fraction. By knowing the expression for the Green function it is easy to find a density of states of the source located in the upper half-space over the surface of hyperbolic metamaterial. Its distance dependance can be derived by using the above mentioned Green function
approach \cite{novotny}
\begin{eqnarray}
\label{rho}
\rho_{HMM}(z,\omega,f)= \frac{2\omega}{\pi c^2} Im \left[tr(G(z,z,\omega,f))\right]
\end{eqnarray}
 Using equation (\ref{polariz}) it can be expressed as a near field enhancement of the  electromagnetic vacuum density of states
 $\rho_0=\frac{\omega^2}{\pi^2 c^3}$
\begin{eqnarray}
\label{rho2}
\rho_{HMM}(z,\omega,f)= \frac{3}{8} \rho_0 \frac{Im \left[R_{12}(\omega,f) \right]}{(kz)^3},
\end{eqnarray}
which becomes large at the small distances from interface of HMM when contribution of the evanescent field is dominant
over its propagating counterpart from vacuum oscillations.
To determine the density of states inside of HMM, we perform standard averaging by replacing delta function in the limit $r\rightarrow 0$ by $1/a^3$,
where $a$ is a characteristic size of the HMM unit cell. Then, density of states becomes
\begin{eqnarray}
\label{rho3}
\tilde{\rho}_{HMM}(\omega,f)= \frac{3}{8} \rho_0 \left(\frac{k_m}{k}\right)^3 Im \left[R_{12}(\omega,f) \right],
\end{eqnarray}
where we use natural wavenumber cut-off $k_m \sim 1/a$ related to the limits of the effective medium approach introduced
in \cite{narim,narim2,narim3}. Here we should note that this density of states would be significantly smaller in value than the one induced
by surface modes (Eq.\ref{rho2}) at distance less than $a$.

\begin{figure}
 \centering
 \includegraphics[width=\columnwidth]{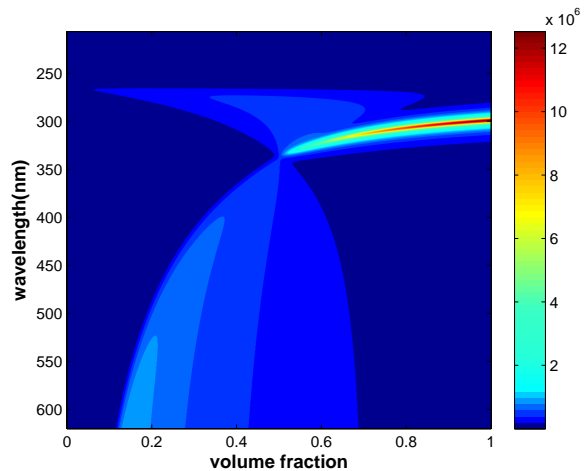}
 \caption{\label{fig:colormap} (color online) A colormap of distribution of density states for HMM with changing of wavelength and fraction of silver inside of the stack.}
 \end{figure}

\begin{figure}
 \centering
 \includegraphics[width=\columnwidth]{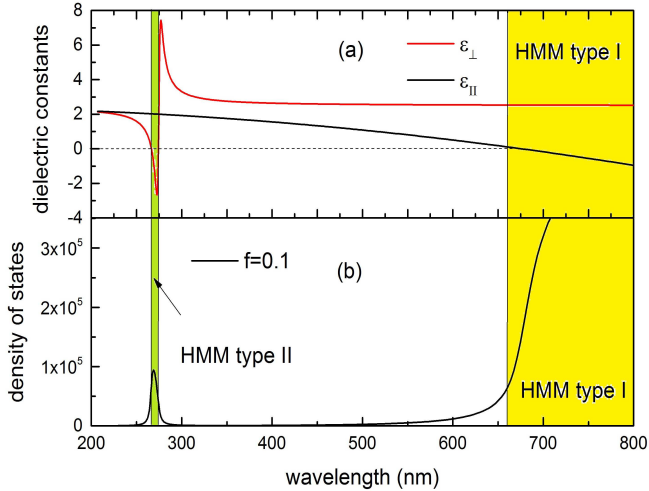}
 \caption{\label{fig:f01} (color online) A demonstration of the photonic's low energy band for HMM of type I with small volume fraction of silver.
 (a) Effective dielectric constant and (b) density of states of HMM}
 \end{figure}

\begin{figure}
 \centering
 \includegraphics[width=\columnwidth]{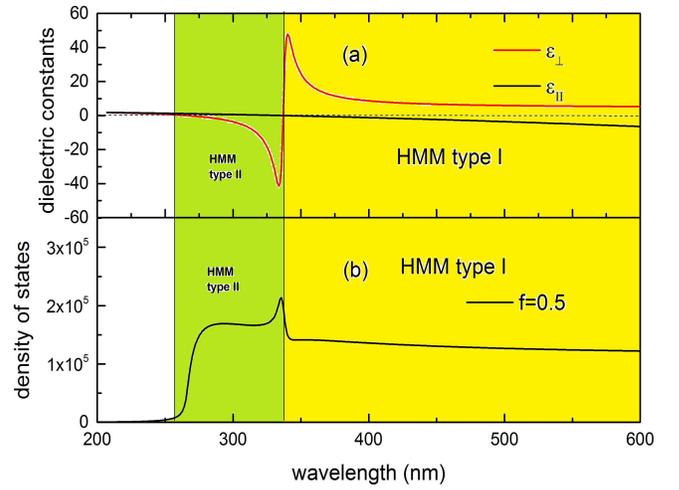}
 \caption{\label{fig:f05} (color online) Photonic energy bands for HMM with volume fraction $f=0.5$ of silver. (a) Effective dielectric
 constants of HMM and (b) density of states of HMM.}
 \end{figure}

To illustrate singular properties of HMM's density of states we have conducted numerical calculations for a multilayer hyperbolic stack with periodically placed silver and silica layers (see Fig.\ref{fig:hmm}(a)).
 \begin{figure}
 \centering
 \includegraphics[width=\columnwidth]{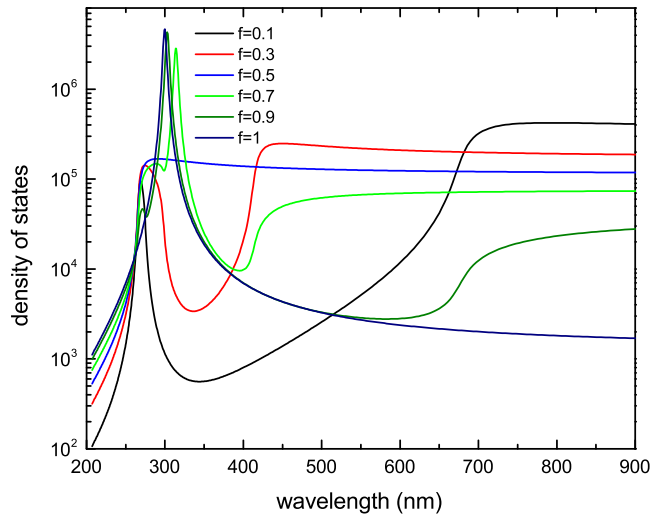}
 \caption{\label{fig:spectral} (color online) A van Hove singularities in the density of states for multilayer HMM with changing
 fraction of silver inside of stack.}
 \end{figure}
%
 A numerical analysis of expression (\ref{rho3})
for this material reveals a set of a van Hove type singularities \cite{hove} in the density of states (see Fig.\ref{fig:spectral}).
 Those singularities are also shown in Figure \ref{fig:colormap} where, as in classical Fermi systems \cite{lifshitz, narim1}, changes from an open to a closed surface are directly related to the singularities in the density of states of an HMM and
 occurs at the saddle point where the derivative of the density of states diverges \cite{bassani}. One may also notice that subsequent increase of the metal fraction leads to formation of the surface plasmon response of HMM for silver resonance region.
%
 \begin{figure}
 \centering
 \includegraphics[width=\columnwidth]{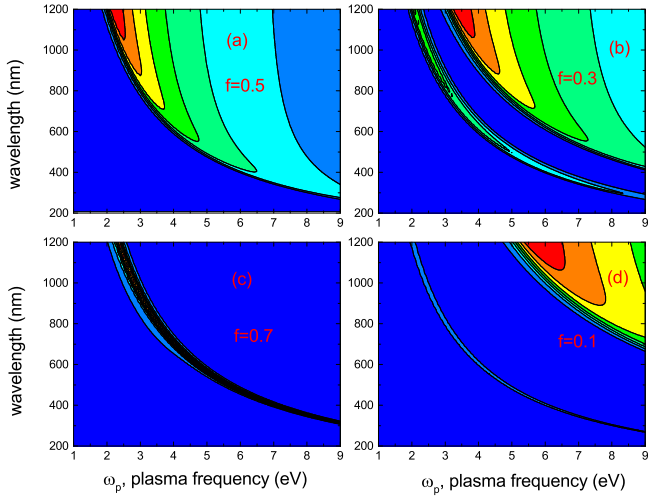}
 \caption{\label{fig:wp} (color online) Demonstration of the photonic band structure of HMM via its density of states with swap over range of plasma frequencies of different materials}
 \end{figure}
While the photonics band structure starts to show up already at the small metal fractions (see, for instance, appearance of the thin bands of HMM type II in Fig.\ref{fig:wp} (b,d) or phase transition between HMM type I and elliptically dispersed medium in Fig.\ref{fig:f01}),
it is interesting to consider situation of direct transition between two states of different hyperbolic dispersions (type I and type II)
that corresponds to touching 3D Dirac point (see Fig.\ref{fig:wp}(a) and Fig.\ref{fig:f05}). In our case, that happens at the spot of the highest symmetry in the system when the volume fraction of silver is $f=0.5$, which appeared to be stable with respect to distortions or imperfections along the interface. One may also relate an observed singularities in the density of states to well-known Fano resonances \cite{fano} that appear at the point of interference of the high k-modes that belong to the both types of HMM.

The striking difference from Fermi systems is the absence of a continuously changing parameter used as a varied difference between chemical potential of the system and its surface energy. For the massless bosonic systems such as light the chemical potential is equal to zero, which
means a gapless state when optical phase transition is passing through the saddle point. The bandgaps can be opened either by breaking P or T symmetries making above mentioned phase transitions topologically non-trivial \cite{lu}. Then density of states would demonstrate really zero values instead of the local minimums presented in this paper. We leave that subject to our forthcoming publications.
%
%
 %

In conclusion, we performed a theoretical and numerical study of the high density of states of hyperbolic metamaterials. We reveal that
the general behavior of HMM photonic density response is reminiscent of Fermi electronic band structure of metal or semiconductor with
local minimum, maximum and saddle point positions defined by photonic bands which are related to the optical transitions between hyperbolic
and elliptical material phases.
 Employing the Green function formalism for multilayer HMM systems, we found a topology protected critical point which appeared to be a saddle point of the system's density of states when the optical phase transition between "neck disrupted" and "neck touched" topologies takes place.

\section{Acknowledgments}
This work was also supported by AFRL Materials and Manufacturing Directorate Applied Metamaterials Program.


\begin{thebibliography}{}
\bibitem{stefano}
O. Di. Stefano, N. Fina, S. Savasta, R. Girlanda and M. Pieruccini,
J.Phys.:Condens.Matter 22, 315303 (2010)
\bibitem{cortes}
C. L. Cortes, W. Newman, S. Molesky and Z. Jacob,
J.Opt. 16, 129502 (2014)
\bibitem{jacob}
Z. Jacob, J.-Y. Kim, G.V. Naik, A. Boltasseva, E.E. Narimanov, V.M. Shalaev
 Appl.Phys. B, 100, 215 (2010)
\bibitem{narim1}
H. Krishnamoorthy, Z.Jacob, E.Narimanov, I.Kretzschmar, V. Menon,
Science, 336, 205 (2012)
\bibitem{narim-ishii}
S. Ishii and E. Narimanov,
Sci. Rep. 5, 17824, (2015)
\bibitem{zhang}
W. Gao, M. Lawrence, B. Yang, F. Liu, F. Fang, B. Béri, J. Li, and S. Zhang
Phys.Rev.Lett., 114, 037402 (2015)
\bibitem{lifshitz}
I.M. Lifshitz,
Sov. Phys. JETP 11, 1130 (1960)
\bibitem{volovik}
G.E. Volovik,
Exotic Lifshitz transitions in topological materials,
 arxiv:cond-mat:1701.06435 (2017)
\bibitem{hove}
L. Van Hove,
Phys. Rev. 89, 1189 (1953)
  \bibitem{shalaev2}
 A. Kildishev, A. Boltasseva, V. Shalaev,
 Science, 339(6125):1232009 (2013)
\bibitem{wein}
C. L. Fefferman, J. P. Lee-Thorp and M. I. Weinstein
PNAS 111 (24) 8759 (2014)
\bibitem{hughe}
See,e.g., B.A.Bernevig, T.L.Hughes, "Topological insulators and topological superconductors", Princeton University Press, Princeton, New Jersey, (2013).
\bibitem{sommer}
A. Sommerfeld,
Ann. Phys. Lpz. 28 665 (1909)
\bibitem{maradudin}
 A. A. Maradudin and D. L. Mills,
 Phys.Rev.B, 11, 1392 (1975)
 \bibitem{markel}
  G. Panasyuk, J. Schotland and V. Markel,
  J. Phys. A: Math. Theor. 42 275203(2009).
 \bibitem{pust1}
 V N. Pustovit, A. M. Urbas, and D. E. Zelmon,
 Phys.Rev.B 94, 235445 (2016)
  \bibitem{sipe}
  O. Kidwai, S. Zhukovsky, and J. Sipe,
  Phys. Rev. B, 85, 053842 (2012).
  \bibitem{sipe2}
  O. Kidwai, S. V. Zhukovsky, and J. E. Sipe,
  Optics Letters, 36, 2530 (2011)  
  \bibitem{maria}
  M. Tschikin, S. Biehs, R. Messina and P. Ben-Abdallah, J. Opt. 15 105101 (2013)  
  \bibitem{shalaev}
  Z. Jacob, J.Y. Kim, G. Naik, A. Boltasseva, E. Narimanov, V. Shalaev,
 Appl Phys B 100, 215 (2010)

 \bibitem{drachev}
V. Drachev, V. Podolskiy, and A. Kildishev,
Optics Express, 21, 15048, (2013)
  \bibitem{christy}
P. B. Johnson and R. W. Christy,
Phys. Rev. B 6, 4370 (1973).
  \bibitem{novotny}
 See,e.g., L. Novotny and B. Hecht, "Principles of Nano-Optics", Cambridge University Press, Cambridge, England, (2012).
  \bibitem{narim}
  J. Zubin,  I. I. Smolyaninov, and E. E. Narimanov,
  Appl. Phys. Lett. 100, 181105 (2012)
  \bibitem{narim2}
 E. Narimanov and I. Smolyaninov,
 arXiv:1109.5444, (2011)
  \bibitem{narim3}
 J.Liu and E.Narimanov,
  Phys.Rev.B 91, 041403 (2015)
  \bibitem{bassani}
 F. Bassani and P. Parravicini, "Electronic States and Optical Transitions in Solids", Pergamon Press. (1975).
\bibitem{hald}
F. D. M. Haldane
Phys. Rev. Lett. 61, (2015)
 \bibitem{fano}
M. F. Limonov, M. V. Rybin, A. N. Poddubny and Y. S. Kivshar
Nature Photonics, 11, 543 (2017)
 \bibitem{lu}
L. Lu, J. D. Joannopoulos and M. Soljacic
Nature photonics, 8, 821 (2014)
\end{thebibliography}
\end{document}